# Reference antenna techniques for canceling radio frequency interference due to moving sources

D. A. Mitchell[1] and J. G. Robertson

School of Physics, University of Sydney, Sydney, New South Wales, Australia

[1]  We investigate characteristics of radio frequency interference (RFI) signals that can affect the excision potential of some interference mitigation algorithms. The techniques considered are those that modify signals from auxiliary reference antennas to model and cancel interference from an astronomical observation. These techniques can be applied in the time domain, where the RFI voltage is modeled and subtracted from the astronomy signal path (adaptive noise canceling), or they can be applied to the autocorrelated and cross-correlated voltage spectra in the frequency domain (postcorrelation canceling). For ideal receivers and a single, statistically stationary interfering signal, both precorrelation and postcorrelation filters can result in complete cancellation of the interference from the observation. The postcorrelation method has the advantage of being applied on tens or hundreds of millisecond timescales rather than tens or hundreds of nanosecond timescales. However, this can be a disadvantage if the RFI transmitter location is changing, since the cross-correlated power measurements which link the interference power in the astronomy and reference signal paths can decorrelate. If the decorrelation is not too severe, it can be allowed for, at the expense of a noise increase. The time domain adaptive cancelers are allowed to slightly vary their internal coefficients and adapt to changing phases during the integrations, which means that they avoid the decorrelation problem. However, the freedom to adapt also results in a noise increase. In this paper the ability of both types of cancelers to excise interference originating from a moving source is compared. The cancelers perform well on both observed and simulated data, giving complete cancellation.

## 1.  Introduction

[2]  No matter what the astronomical application, radio frequency interference (RFI) is becoming an increasing problem in radio astronomy, and many methods for removing or suppressing the RFI are being proposed, evaluated and implemented. In most cases, the astronomer's targets are the correlations of signals from one or more antennas, and it is only these longer-term time averages of power that are wanted—there is no requirement to preserve voltage modulation. These applications are generally either finding the autocorrelation of signals from a single antenna (to measure the power spectrum of the astronomy signal), or the cross correlation of signals from more than one antenna (to measure the spatial coherence—or visibilities—of the astronomy signal). For details and specific examples see *Thompson et al.* [1986] and various chapters of *Taylor et al.* [1999].

[3]  Suppose that a sampled voltage stream consists of an additive mixture of components that are uncorrelated with each other. Suppose also that each component is statistically stationary so that if the component happens to be present in more than one voltage stream the phase difference and the ratio of the sampled amplitudes measured at two receivers are constant. Then as de-

---

[1]Also at Australia Telescope National Facility, Commonwealth Scientific and Industrial Research Organisation, Epping, New South Wales, Australia.

scribed by *Briggs et al.* [2000], if one of these correlated components is undesired and hindering our ability to probe cosmic components, it is possible to cancel this RFI from the power spectra, after the voltages have been correlated. Canceling RFI from correlations, referred to as "postcorrelation canceling," can offer many advantages over canceling the additive RFI voltage directly, particularly in regard to computational efficiency, since the canceling is performed on each correlation, tens or hundreds of times a second, rather than each voltage sample, tens or hundreds of times a millisecond. Furthermore, it is possible to implement the technique in some current arrays with no modification, albeit at reduced array performance, for example [*Kesteven*, 2002]. Before considering cancellation techniques further, the signal itself needs to be briefly described.

[4] Consider the voltage sequence sampled at an antenna as a combination of three complex additive components: receiver noise, $N(\nu, t)$; a noise-like cosmic component, $S(\nu, t)$; and interference, $I(\nu, t)$. Since each quasi-monochromatic spectral channel of the signal will be considered independently, the frequency labeling will be dropped to condense equations. After experiencing a phase shift, $\phi_m(t)$, due to the geometric delay of the signal relative to an arbitrary reference point, and being amplified and possibly phase shifted by a gain term, $G(t)$, the signal at the output of antenna $m$'s sampler at time step $i$, which is only measuring the voltage in the narrow spectral band centered at frequency $\nu$, is

$$V_m(i) = S(i) + G_m(i)I(i)e^{j\phi_m(i)} + N_m(i), \qquad (1)$$

where it is assumed that the signal has been amplified and delayed so that the cosmic signal is in phase with equal power at all of the receivers.

[5] In the absence of the RFI component, one could detect and measure the amount of cosmic power by comparing the voltage sequences from two antennas, $V_l$ and $V_m$, since the background noise is different for the two receivers. However, the presence of the interfering signal obscures the cosmic detection. If one were to compare each main antenna voltage sequence with voltage sequences from antennas that do not measure the cosmic signal (so only comparing the RFI components of various signals), then one could attempt to model and remove the RFI component in the comparison of $V_l$ and $V_m$ and recover the astronomy signal. In practice reference antennas, such as parabolic reflectors pointed in the direction of a known transmitter, will measure some cosmic power through their sidelobes, but it is assumed that this is negligible compared with the receiver noise power. Assuming a negligible cosmic contribution, the reference signals are of the form

$$V_r(i) = G_r(i)I(i)e^{j\phi_r(i)} + N_r(i). \qquad (2)$$

[6] For any pair of antennas, the signals are compared by correlating the two voltage sequences together, that is, multiplying one signal by the complex conjugate of the other and accumulating the product for some accumulation time (~1 s). Uncorrelated components will multiply to give zero mean noise which will average away as the number of samples accumulated approaches infinity, while a component that is present in both signals will correlate constructively, with an amplitude proportional to its power.

[7] The correlated power terms in the obscured main antenna cross correlation (so ignoring zero mean noise terms) are

$$\begin{aligned} P_{lm} &= \langle V_l(i)V_m^*(i) \rangle \\ &= \sigma_S^2 + G_l G_m^* e^{j\phi_{lm}}\sigma_I^2 + \langle N_l(i)N_m^*(i) \rangle, \end{aligned} \qquad (3)$$

where $\phi_{lm} = \phi_l - \phi_m$, $\sigma_S^2$ and $\sigma_I^2$ are the variances of the cosmic and interfering signals respectively, the asterisk superscript indicates a complex conjugation, the angular brackets represent the expectation operator which is approximated by a time average, and it has been assumed that the gain and phase terms are constant over the time interval. The cross-correlated receiver noise term, which should be zero mean, has been included to remain general, since the following techniques also apply to autocorrelations where the noise is correlated against itself, that is, when $l = m$.

[8] Postcorrelation cancelers, which are briefly reviewed in the following section, estimate and then subtract the RFI power, $G_l G_m^* e^{j\phi_{lm}}\sigma_I^2$, from $P_{lm}$.

## 2. Postcorrelation Cancelers

[9] The postcorrelation technique described by *Briggs et al.* [2000] involves creating a model of the RFI in the main astronomy correlations using signals from a set of auxiliary reference antennas. Keeping in mind that each spectral channel is processed separately, the RFI model is a complex number with an amplitude and phase equal to the RFI component in (3).

[10] Following *Briggs et al.* [2000], the postcorrelation canceler estimates $G_l G_m^* e^{j\phi_{lm}}\sigma_I^2$ in $P_{lm}$ using the closure relations, resulting in an RFI power model

$$\begin{aligned} M_{lm} &= \frac{P_{lr_1}P_{mr_2}^*}{P_{r_1 r_2}^*} \\ &= \frac{\langle G_l G_{r_1}^*\sigma_I^2 e^{j\phi_{lr_1}} \rangle \langle G_m G_{r_2}^*\sigma_I^2 e^{j\phi_{mr_2}} \rangle^*}{\langle N_{r_1}N_{r_2}^* \rangle^* + \langle G_{r_1} G_{r_2}^*\sigma_I^2 e^{j\phi_{r_1 r_2}} \rangle^*} \\ &\approx G_l G_m^* e^{j\phi_{lm}}\sigma_I^2, \end{aligned} \qquad (4)$$

which is equal to the RFI term in $P_{lm}$.

[11] In practice the expectation operators are not infinite in extent, and there is zero mean noise centered on the correlated power in (4). So when $M_{lm}$ is subtracted from $P_{lm}$ the residual power will be zero mean noise, not zero, and there will always be an increase in noise over the situation where there was no RFI to begin with. This noise will average toward zero, and essentially result in an increase in system temperature (a decrease in the sensitivity of $P_{lm}$).

[12] The closure relations which suggest the equality of the RFI in $P_{lm}$ and the model $M_{lm}$ only hold when the phase difference of the RFI signal remains constant for all antenna pairs. In practice though the geometric delays of the RFI signal between various antennas are not constant, because of the apparent motion of the RFI transmitter with respect to the array, which results in decorrelation of the cross-power measurements in (3) and (4) [see, e.g., *Thompson et al.*, 1986]. The relative antenna gain to the RFI for each signal will also vary, but this is not considered here. If the only broken assumption is that of variable RFI geometric delays, but the delays are essentially constant over the time average, canceling before or after correlation will give very similar results. However, when the delays are changing appreciably during the integration, the different timescales on which the precorrelation and postcorrelation algorithms are applied lead to differences in the canceling. To investigate the differences, we will first look at the effect of variable delays on the postcorrelation algorithm, then precorrelation cancelers which can track the changing delays during the time integrations.

## 3. Decorrelated Cross Power

[13] If the RFI signal geometric delays used in (3) or (4) are changing while the postcorrelation RFI model is being calculated, the RFI power will be smeared over a range of delays, and the RFI model will be incorrect. This process, occurring because of the apparent motion of the array as it tracks cosmic sources—or the motion of the RFI source itself—is known as time average smearing and the resulting loss of correlated power is known as fringe rotation decorrelation, since the source is moving through fringes. If, at time $t_0$, the gain terms stay approximately constant over the correlation averaging time, $T$ s, then the correlated RFI power for arbitrary signals $V_j$ and $V_k$, written now as a function of delay rather than phase, is

$$P_{I,jk} = \frac{G_j G_k^*}{T} \int_{t_0-\frac{T}{2}}^{t_0+\frac{T}{2}} e^{j2\pi\nu\tau_{jk}(t)} dt. \quad (5)$$

[14] Let the geometric delay for the source at time $t_0$ be $\tau_{jk_0}$ s. If the rate of change of $\tau_{jk}$ with time is approximately constant over the integration and equal to $\Delta\tau_{jk}/T$ s of delay per second, the chain rule can be used to write the measured RFI power in (5) as a function of the change in geometric delay over the course of the integration

$$P_{I,jk}(\Delta\tau_{jk}) = \frac{G_j G_k^*}{\Delta\tau_{jk}} \int_{\tau_{jk_0}-\frac{\Delta\tau_{jk}}{2}}^{\tau_{jk_0}+\frac{\Delta\tau_{jk}}{2}} e^{j2\pi\nu\tau_{jk}} d\tau_{jk}$$
$$= \mathrm{sinc}(\nu\Delta\tau_{jk})P_{I,jk}(0), \quad (6)$$

where $\mathrm{sinc}(x) = \sin(\pi x)/(\pi x)$. The faster the delay is changing (and the higher the frequency), the more the cross-power estimate will be smeared out. The proportion of correlator output remaining after decorrelation is the ratio of $P_{I,jk}(\Delta\tau_{jk})$ to $P_{I,jk}(0)$, and for baseline $jk$ is denoted $F_{jk}$. For the constantly varying geometric delay of $\Delta\tau_{jk}$ s per integration, we have

$$F_{jk} = \mathrm{sinc}(\nu_0 \Delta\tau_{jk}). \quad (7)$$

[15] The effect of decorrelation can be seen in Figure 1. This data is Global Positioning System (GPS) satellite interference collected at two Australia Telescope Compact Array antennas separated by 4.4 km. The RF voltages were centered at 1575 MHz in a 4 MHz wide band, sampled with 4 bits, and summarized by *Bell et al.* [2001]. Figure 1a is a plot of cross-correlated power (scaled to be 1 for no decorrelation) for various integration lengths, which fits well to the theoretical curve given by (7) as the satellite moved across the sky.

[16] So for RFI with an apparent motion relative to the sky reference frame, the postcorrelation model for signals $V_l$ and $V_m$ given in (4) becomes

$$M_{lm} = \frac{F_{lr_1} F_{mr_2}}{F_{r_1 r_2}} G_l G_m^* \sigma_I^2, \quad (8)$$

while the RFI power in the main cross-power measurement will have decorrelated to $F_{lm} G_l G_m^* \sigma_I^2$, leaving residual power of

$$R_{lm} = \left(F_{lm} - \frac{F_{lr_1} F_{mr_2}}{F_{r_1 r_2}}\right) G_l G_m^* \sigma_I^2. \quad (9)$$

[17] Figure 1b shows the proportion of GPS satellite power remaining in the power spectrum of one ATCA antenna ($l = m$) after cancellation using both polarizations from the other ATCA antenna as the two references (one would typically not choose to place reference antennas 4.4 km from the main antennas, this is a worst case example). Since the references are collocated, the only $F_{jk}$ terms that will not be equal to 1 in (9) are $F_{mr_1}$ and $F_{mr_2}$ (written $F_{mr}$ since they are equal). It is clear that

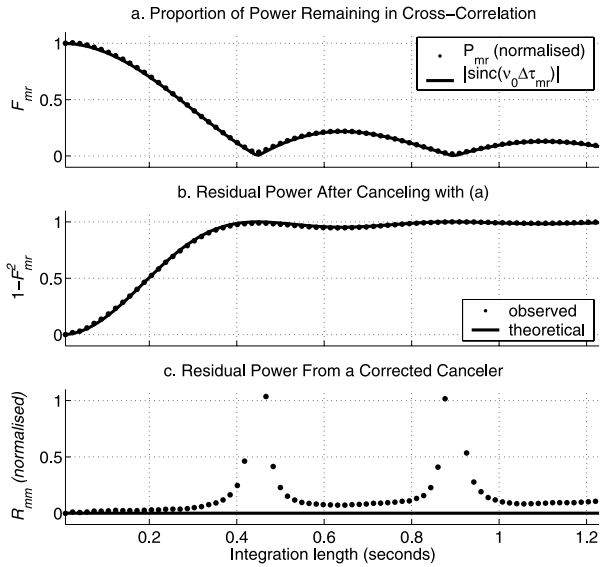

**Figure 1.** (a) Proportion, $F_{mr}$, of the GPS power remaining after integration as a function of integration time $T$. Dots show the measured power, and the line shows the theoretical estimate. (b) Proportion of residual RFI remaining in the output power after cancellation with $M_{mm}$. (c) Proportion of residual RFI remaining in the output power after cancellation using the modified model, $M'_{mm}$.

the cancellation becomes ineffective when decorrelation noticeably affects the cross-power measurement.

[18] From equations (8) and (9) it can be seen that one can modify the postcorrelation algorithm so that it takes the decorrelation into account

$$M'_{lm} = \frac{F_{lm}F_{r_1 r_2}}{F_{lr_1}F_{mr_2}} M_{lm}, \quad (10)$$

where the prime on the model indicates that fringe rotation has been corrected for. This is shown in Figure 1c. Apart from the two regions where $F_{mr}$ goes to zero and the noise is being infinitely magnified, the canceler achieves complete cancellation of the RFI, albeit with a small magnification of the noise. In those regions where most of the RFI power has been decorrelated away, so that one or more of the $F_{jk}$ terms is close to zero, there is little that can be done to reverse the decorrelation. If it happens in the cross-power spectrum case that most of the RFI power has been decorrelated out of $P_{lm}$, then it might be better to do nothing as the RFI is canceling itself.

[19] Figure 2a shows the residual spectra for three of the integration lengths from Figure 1 when decorrelated RFI models, $M_{mm}$, were used. Figure 2b shows the same

spectra when modified models, $M'_{mm}$, were used. The RFI is the peak centered at spectral channel 200, and the improvement achieved by scaling the weights is clear.

[20] We will now show how, with constant geometric delays, equivalent canceling can be performed on the RFI voltages directly. This type of canceling needs to be applied much more often, however as discussed afterward, it has the ability to track the changing delays.

## 4. Voltage Cancelers

[21] Single reference precorrelation adaptive noise cancelers are discussed by *Widrow and Stearns* [1985] and an early application to radio astronomy is described by *Barnbaum and Bradley* [1998]. A precorrelation canceler which uses two reference signals to achieve equivalent cancellation to the postcorrelation canceler is shown schematically in Figure 3. These dual-reference filters were suggested by *Briggs et al.* [2000] and are discussed by *Mitchell et al.* [2002]. Each spectral component of one reference signal is amplified and phase shifted by a weight so that its RFI component matches the RFI component in the main signal. A second reference is used to determine the required weight, which is chosen to set the correlation between the reference signal and the canceler output to zero. The second reference has receiver noise that is uncorrelated with the first, so that RFI is the only signal that could be in both the second reference voltage and the output voltage. Thus zeroing

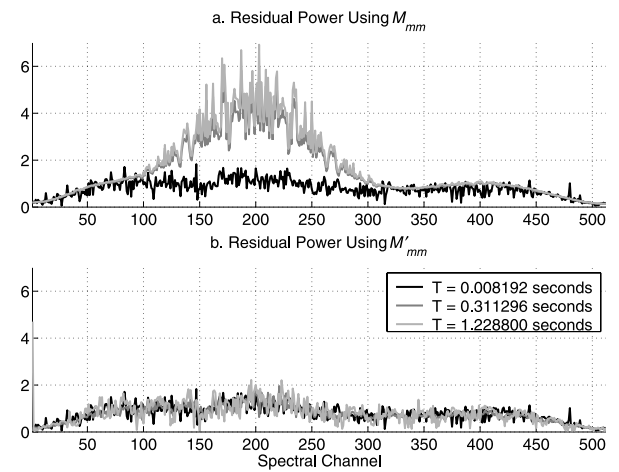

**Figure 2.** Residual power spectra after three integrations of length 0.008, 0.311, and 1.229 s for (a) the standard post-correlation filter, $M_{mm}$, and (b) the modified post-correlation filter, $M'_{mm}$.

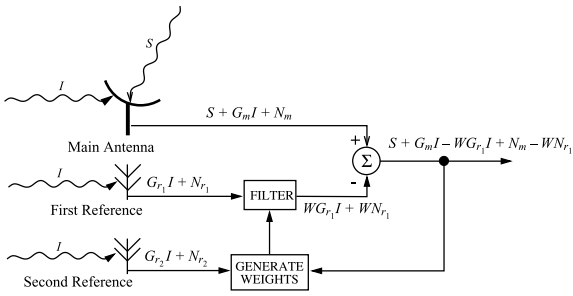

**Figure 3.** Adaptive canceler. The complex weights are those which set the cross correlation between the output and the second reference spectrum to zero.

the cross correlation means zeroing the RFI in the output. From Figure 3, the weight which sets the correlated power of the canceler output and reference $r_2$ to zero, $\langle (V_m - W_{mr_2} V_{r_1}) V_{r_2}^* \rangle = 0$, is

$$W_{mr_2} = \frac{\langle V_m V_{r_2}^* \rangle}{\langle V_{r_1} V_{r_2}^* \rangle}, \qquad (11)$$

which, as in the postcorrelation case, will be incorrect if the expectation values are estimated from cross-power measurements that have decorrelated because of changing RFI delays. The weights, which are optimal at any given instant, can be tracked rather than estimated statistically, as described by *Widrow and Stearns* [1985] and discussed in section 5. For the moment assume that we know the expectation values in (11). Even though a RFI voltage model can be made to completely cancel the RFI in $V_m$, there will still be power in the output because of reference receiver noise (it is this noise which prompted the use of the second reference). To insure that the receiver noise added during the filtering of $V_m$ will not be correlated with the receiver noise added during the filtering of $V_l$, we need to weight a different reference signal when main signal $V_l$ is processed. Since the added receiver noise is all from the first reference receiver, swapping the references around for $V_l$ means that noise from the second reference receiver will be added, as illustrated in Figure 4. The weight for $V_l$ is

$$W_{lr_1} = \frac{\langle V_l V_{r_1}^* \rangle}{\langle V_{r_2} V_{r_1}^* \rangle}. \qquad (12)$$

[22] When $W_{mr_2}$ and $W_{lr_1}$ are used to filter $V_m$ and $V_l$ respectively, the output cross-power measurement is

$$P_{lm} = \langle (V_m - W_{mr_2} V_{r_1})(V_l - W_{lr_1} V_{r_2})^* \rangle. \qquad (13)$$

[23] As in the postcorrelation case, this has completely removed the RFI but added zero mean noise which decreases the sensitivity of the cross-power measurement. However, the canceler does not leave a correlated signal which would set a maximum achievable sensitivity level.

[24] While the postcorrelation algorithm described earlier can only account for the decorrelation after it has occurred, we now look at how the precorrelation algorithm can be made to track any changing delays (one could of course account for the decorrelation in the precorrelation weights by altering (11) and (12), with a result similar to that of postcorrelation cancelers).

## 5. Adapting Time Domain Cancelers

[25] When the optimal precorrelation filter weights given by the expectation values in (11) and (12) are changing in time because of varying geometric delays, adaptive cancelers can attempt to keep up with the changing delays, by slightly modifying their weights at a rate equal to twice the bandwidth for Nyquist sampling [*Widrow and Stearns*, 1985].

[26] While the equations for RFI models given in section 4 give a single complex number for each spectral channel, the precorrelation cancelers were actually applied in the time domain, directly to the voltages. In this situation, a set of reference voltage samples, delayed in time in both the positive and negative direction, are amplified by a real valued gain factor (the model weights), and then the weighted reference samples are added together to form the RFI model for the current sample. It should be pointed out that the time and frequency domain cancelers are essentially equivalent.

[27] An example weight vector (refer to the simulation in the next section) is shown in Figure 5. This RFI is wideband noise spread over the entire spectrum with equal power, so a suitable weight vector would be single peak at an appropriate delay (the negative of the geometric delay between the main and reference antennas) with an amplitude equal to the value the reference signal needs to be multiplied by to match the RFI in the main signal. This is indeed the weight vector that the algorithm gave—except for a small zero mean noise floor (this "misadjustment" noise, due to the fact that the weights are being adapted using noisy power measurements, is discussed at length by *Widrow and Stearns* [1985].

[28] Given a set of starting weights and the associated output power, *Widrow and Stearns* [1985] give an estimate for the correct weights using the local power gradient with respect to weights. For time step $i + 1$, the offset of the weight from the optimal value for

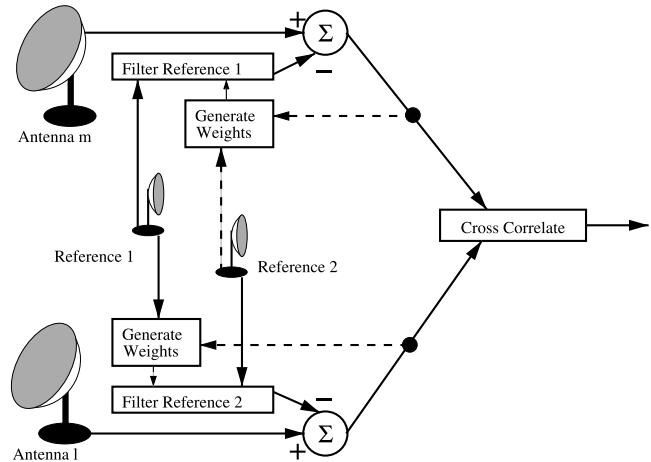

**Figure 4.** Double canceler setup where the model for each main signal is generated from a unique reference.

the $k^{th}$ delayed reference sample can be found from time step $i$

$$w_{i+1}(k) = w_i(k) - \frac{1}{2\sigma_r^2}\Delta_i(k), \quad (14)$$

where $\sigma_r^2$ is the variance of the reference signal and $\Delta_i(k)$ is the gradient of the canceler output power with respect to the $k^{th}$ weight at the previous time step. Instead of attempting to find the optimal weights in one step (which would require using statistics as in section 4), the one step procedure can be replaced by an iterative process in which the weights are only slightly modified at each iteration, $w_{i+1}(k) = w_i(k) - \mu\Delta_i(k)$, where $\mu$ is a user defined constant called the adaptation gain factor which controls the rate of adaptation. *Widrow and Stearns* [1985] derive the equation for the power gradient term and show the weight adaptation to be equal to

$$w_{i+1}(k) = w_i(k) + 2\mu v_{out,m}(i)v_r(i-k), \quad (15)$$

where $v_{out,m}$ is the voltage at the output of the canceler for antenna $m$, and $\mu$ must be between 0 and $\frac{1}{(L+1)\sigma_r^2}$ for convergence to hold, where $L + 1$ is the number of delayed reference voltage samples used. Equation (15) shows that the iterative alteration of the weights really does form the correlation between the reference and the output and adapts until this correlation is zero. The value of $\mu$ controls the length of the integration, smaller values will smooth (reduce) the noise fluctuations in the weights, but increase the time constant for their response to changes (such as in the geometric delay).

[29] Adaptive time domain cancelers can follow a moving source, avoiding decorrelation, but will eventually suffer a noise penalty when they have to adapt very quickly.

## 6. Simulated Comparisons

[30] In this section the ideas discussed up to this point will be investigated using simulated wideband Gaussian noise. The algorithms work equally well for signals that are limited to only part of the frequency band, however a stabilizing threshold may be required when generating the weights [*Mitchell et al.*, 2002], and if the signal is changing in frequency the rate of adaptation must be fast enough to track the weight variation (just as the weights must be able to track a changing delay). To keep things simple and minimize the number of correlations affected by the changing delays (as in the GPS data from

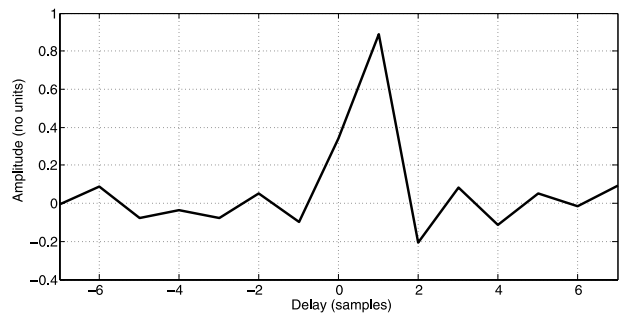

**Figure 5.** Example weight vector. The reference vector is delayed by each of the delays along the horizontal axis and is multiplied by the corresponding amplitude, then all of the weighted copies are added together to give the model voltage estimate.

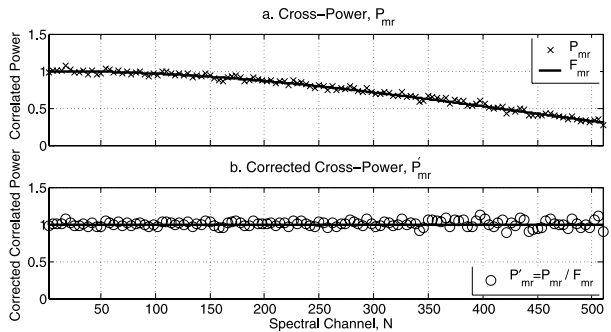

**Figure 6.** (a) Normalized cross-correlated RFI power, $P_{mr}$, on the baseline between the main and reference antennas which has been decorrelated. (b) Cross-correlated power after allowing for decorrelation, $P'_{mr}$.

section 3), consider the situation of a single antenna observing a cosmic source in the presence of wideband RFI. The quantity we wish to measure is the cosmic source power, which can be achieved by calculating the autocorrelation of the received voltage sequence, but this will also contain receiver noise and RFI power. Power measurements are made while implementing both precorrelation and postcorrelation cancellation algorithms, both of which will be affected by the changing delays, as discussed in the previous sections. Since an autocorrelation is being considered, the signal index, $l$, becomes $m$ in all of the previous equations and the RFI in the main signal autocorrelation will suffer no decorrelation, $F_{mm} = 1$. Also, consider two reference signals coming from orthogonal polarizations of a single reference antenna (antenna $r$), so that the references are collocated and the RFI in their cross correlation will also suffer no decorrelation. So long as the RFI is 100% polarized (and not orthogonal to the polarization of any of the receivers), then $F_{rr} = 1$, and the two cross-power measurements between signal $m$ and each of the reference signals are decorrelated down to a fraction $F_{mr}$ of the incident power. Equations (8), (9) and (10) reduce to

$$M_{mm} = F_{mr}^2 G_m G_m^* \sigma_I^2$$
$$R_{mm} = (1 - F_{mr}^2) G_m G_m^* \sigma_I^2 \quad (16)$$
$$M'_{mm} = \frac{1}{F_{mr}^2} M_{mm} = P_{I,mm}.$$

[31] To investigate these equations, three voltages signals were generated, $V_m$, $V_{r_1}$ and $V_{r_2}$, each of which was broadband Gaussian noise. The main voltage was the sum of a broadband Gaussian RFI sequence and a unique Gaussian noise sequence simulating receiver noise. The reference signals contained a delayed and amplified version of the RFI and their own unique receiver noise sequences. As the voltages were generated the delay between the RFI in the main voltage sequence and the RFI in the reference voltage sequences was slightly incremented, simulating a moving source.

[32] Figure 6 shows RFI power in the cross correlation of signal $m$ and one of the references, along with the RFI power after the decorrelation has been corrected for. The decorrelated spectrum, $P_{mr}$ in Figure 6a, is divided by $F_{mr}$ (the line), to give $P'_{mr}$, shown in Figure 6b. The power level in each frequency channel has been correctly reset to be centered on the expected (decorrelation free) value. The noise in the power measurement has increased, which is particularly clear in the higher frequency channels where more amplification was needed.

[33] Figure 7 is a gray scale map of the adaptive precorrelation weight vector (refer to Figure 5) displayed vertically as the integration progresses through the voltage samples. In the simulations $L = 16$, that is, 8 negative and 8 positive delays. The adaptation gain factor was set to be 1 percent of the delay change over the integration (1% of 1.5 Nyquist samples), multiplied by the maximum value allowed; $\mu = \frac{1.5}{100} \frac{1}{(L+1)\sigma_r^2}$. The geometric delay has changed by 1.5 sample lengths over the course of the integration. The gray value of each pixel indicates the amplitude of the weight, and as the delay changes, the main peak in the weight vector follows it.

[34] The residual power after canceling is shown in Figure 8. The dashed line shows the theoretical level of reference receiver noise added during filtering when the RFI transmitter is stationary and the geometric delay is constant; zero mean noise which will continue to average toward zero as the integration length is increased. The adaptive filter results (indicated by the dots), follow this theoretical limit across the entire spectrum, indicating that the filter is indeed tracking the changing delays. The solid line and crosses represent the postcorrelation canceler which has not been modified to account for decorrelation. The excess power above the dotted line is residual interference which will not average down any further. The residual power after postcorrelation canceling using $M'_{mm}$, which has accounted for fringe rotation (Figure 6b), is indicated by open circles and continues to average down toward the theoretical limit as the frequency channel and decorrelation increase. The residual power in the adaptive and corrected postcorrelation filtered spectra does not contain RFI, and will average down with the dashed light blue line, albeit with a larger noise floor when more fringe rotation correction is needed (that is, the higher frequency channels in Figure 6).

[35] So while the postcorrelation canceler cannot be made to track an interfering signal's geometric delay faster than the integration length, the resulting decorre-

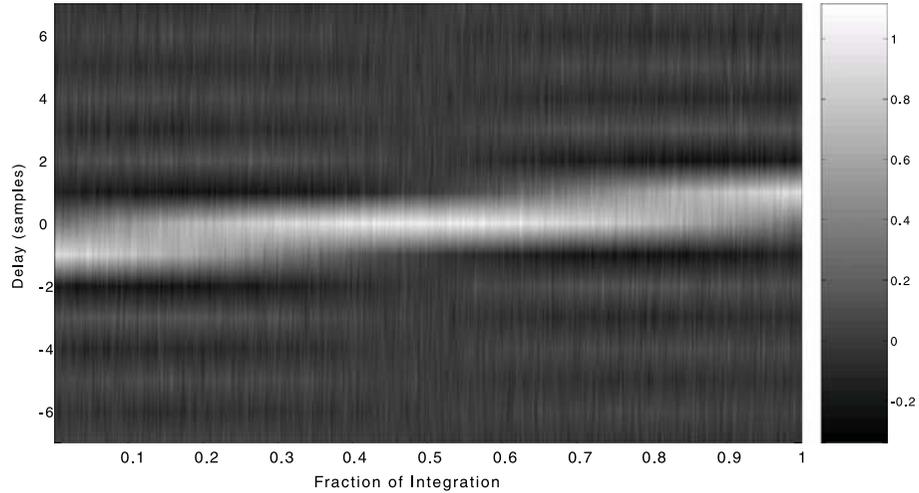

**Figure 7.** Weight vectors as a function of integration number as the integration progresses. Gray scale indicates the amplitude of the weights.

lation of the cross correlations in the filter equations can often be accounted for.

## 7. Conclusions

[36] Both the precorrelation and postcorrelation RFI cancelers discussed in this paper are adversely affected by interfering sources that are moving relative to cosmic sources. This is due to changing geometric delays causing decorrelation of the cross-power measurements used to generate the RFI models. If the rate of change of the geometric delays is known and the amount of decorrelation is only slight, the models can be corrected by amplifying the cross-power measurements back to their theoretical levels. This process can be avoided in precorrelation cancelers, since they can track the correct models much faster (closer to the voltage sample rate which is to the order of the signal bandwidth). Both of these filters result in extra noise relative to stationary RFI situation.

[37] One should keep in mind the particular case of terrestrial interference from a fixed source. In this situ-

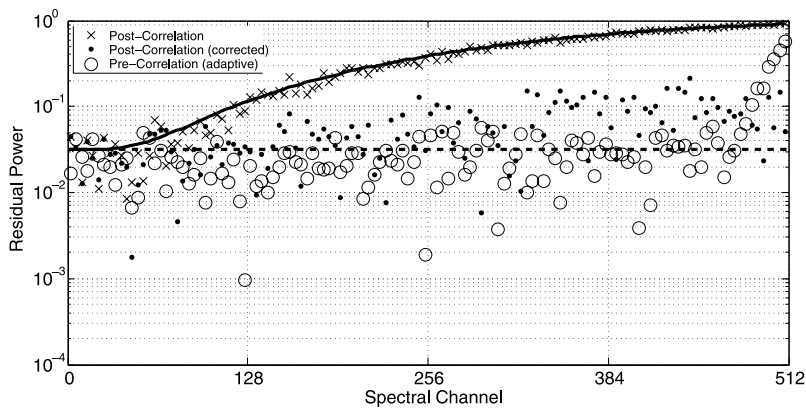

**Figure 8.** Residual power after filtering a wideband RFI signal (normalized to unity for no filtering). The lines indicate the theoretical amount of residual power remaining: The solid line is for a moving source, and the dashed line is for a stationary source. Points indicate the simulation results: Dots are for the adaptive precorrelation canceler, and crosses and open circles are for postcorrelation canceler with and without decorrelation corrections, respectively.

ation the interferometer is fringe tracking the celestial sphere so the RFI fringe rate is the negative of the sidereal fringe rate. Since the correlator is inserting this phase term, one always knows the fringe rate for each pair of antennas, and for antenna pair $jk$, can calculate $F_{jk}$ exactly. The required level of suppression and added complexity, however, may dictate that shorter integration times are necessary to eliminate the need for the decorrelation corrections.

[38] **Acknowledgment.** We acknowledge Mike Kesteven, Frank Briggs, Bob Sault, and Lawrence Cram for many enlightening discussions on the topic and the Scientific Committee on Frequency Allocations for Radio Astronomy and Space Science (IUCAF) and the University of Sydney for funding to attend the Workshop in Mitigation of Radio Frequency Interference in Radio Astronomy.

———

D. A. Mitchell and J. G. Robertson, School of Physics, University of Sydney, Bldg A28, Sydney, NSW 2006, Australia. (mitch@physics.usyd.edu.au; jgr@physics.usyd.edu.au)